\documentclass[fleqn,12pt,twoside]{article}
\usepackage{espcrc1}

\usepackage{graphicx}
\usepackage[figuresright]{rotating}

\newcommand{\AmS}{{\protect\the\textfont2
  A\kern-.1667em\lower.5ex\hbox{M}\kern-.125emS}}

\title{Particle density fluctuations}

\author{Bedangadas Mohanty for the  WA98 Collaboration \footnote{ 
Presented at Quark Matter 2002, Nantes, France. email : bmohanty@veccal.ernet.in } \\
\vspace{0.3cm}
{\footnotesize
M.M.~Aggarwal$^{a}$, 
Z.~Ahammed$^{c}$,
A.L.S.~Angelis$^{d}$, 
V.~Antonenko$^{e}$, 
V.~Arefiev$^{f}$, 
V.~Astakhov$^{f}$,
V.~Avdeitchikov$^{f}$, 
T.C.~Awes$^{g}$, 
P.V.K.S.~Baba$^{h}$, 
S.K.~Badyal$^{h}$
% ...
\\
}
\vspace{0.3cm}
\noindent
$^{a}$Univ. of Panjab (India),
% ...
$^{c}$VECC, Calcutta (India),
$^{d}$Univ. of Geneva (Switzerland),
$^{e}$Kurchatov (Russia),
$^{f}$JINR, Dubna (Russia),
$^{g}$ORNL, Oak Ridge (USA),
$^{h}$Univ. of Jammu (India),
% ...
}      
\begin{document}

\maketitle

\begin{abstract}
Event-by-event fluctuations in the multiplicities of charged particles 
and photons at SPS energies are discussed. Fluctuations are studied by 
controlling the centrality of the reaction and rapidity acceptance of 
the detectors. Results are also presented 
on the event-by-event study of correlations 
between the multiplicity of charged particles and photons to search for 
DCC-like signals. 
\end{abstract}

\section{Introduction}

At high temperatures or high baryon number density, Quantum 
Chromo-Dynamics (QCD) describes a world of weakly interacting quarks and 
gluons very different from the hadronic world in which we live. This suggests 
the possibility of a phase transition as the temperature or density is 
increased in which there is a transition from a state of matter where 
quarks are confined inside hadrons to one where quarks are free to move
around ({\it deconfined} ) within a large volume - the 
{\it quark gluon plasma} (QGP). This can be 
addressed through experimental studies involving relativistic heavy-ion 
collisions. Experimental searches have focused on isolating signatures 
of two types of phase transitions which might occur in extremely hot and/or 
dense nuclear matter. One is related to the deconfinement of quarks while 
the other is related to chiral symmetry restoration.  There is also an 
interesting possibility of the existence of a {\it tri-critical} point in the 
phase diagram ~\cite{tricritical}, where the transition changes from 
first to second order. 
At the tri-critical point, one would observe singularities in several 
thermodynamical variables, such as the 
specific heat and matter compressibility. 
The thermodynamical variables are related to event-by-event fluctuations in 
experimental observables like particle multiplicity, transverse energy, and 
mean transverse momentum. For example, the total heat capacity is related to 
transverse momentum fluctuations and matter compressibility to multiplicity 
fluctuations~\cite{fluc_thermo,henning}. 
Hence the existence of the tri-critical point can be probed in an 
experiment by varying the control parameters of the reaction 
(impact parameter, rapidity acceptance, beam energy, and system size) 
and studying the event-by-event fluctuations in the global observables. 
It is also believed that in high energy heavy-ion collisions there is a
possibility of creating a chiral symmetry restored phase.
One of the possible interesting consequences of chiral symmetry restoration, 
is the formation of {\it disoriented chiral condensates} (DCC)~\cite{dcc}.
The detection and study of a DCC state is expected to provide valuable
information about the chiral phase transition and vacuum structure of
strong interactions. The probability distribution
of the neutral pion fraction in a DCC domain follows the relation~:
\begin{equation}
P(f) = 1/2\sqrt{f} ~~~~~~~~~~~~~~~~{\rm where}~~~~~~~~~~
 f = N_{\pi^0}/N_{\pi}.
\end{equation}
Since the majority of charged particles consist
of charged pions and majority of photons originate from $\pi^0$
decays, DCC formation in a given domain would be associated with
large correlated event-by-event fluctuations in the multiplicities of
charged particles and photons. Here we present the experimental
results of multiplicity fluctuations
and correlations in relativistic heavy-ion (Pb+Pb) collisions as measured by 
the WA98 experiment~\cite{wa98} at the CERN SPS.

\section{Multiplicity fluctuations }

Recently, much theoretical interest has been generated on the subject of 
event-by-event fluctuations, primarily motivated by the nearly perfect 
Gaussian distributions of several observables 
($N_{\gamma}$, $N_{ch}$, $E_{T}$, and $< p_T > $) for a fixed centrality 
bin~\cite{henning,WA98-15}.  One may define the relative fluctuation as,
\begin{equation}
\omega_X = \frac{\sigma_X^2}{<X>}
\end{equation}
where $\sigma_X^2$ is the variance of the distribution.
The fluctuation is then studied by varying the control parameters of the 
reaction, such as the centrality and rapidity acceptance. 

The WA98 experiment at the CERN SPS
has carried out a detailed analysis of photon and charged
particle multiplicity~\cite{WA98-15} fluctuations. 
The photon multiplicity is measured
using the photon multiplicity detector (PMD) ~\cite{WA98-9} which has the 
pseudo-rapidity coverage from 2.9 to 4.2. The charged particle
multiplicity is obtained from the silicon pad multiplicity detector
(SPMD)~\cite{WA98-3} with pseudo-rapidity coverage from 2.35 to 3.75. 
The centrality of the reaction is defined through the measurement of 
transverse energy ($E_{T}$) with the mid-rapidity calorimeter (MIRAC) having 
a pseudo-rapidity coverage from 3.5 to 5.5.

\subsection{Centrality dependence of multiplicity fluctuations }

\begin{figure}
\begin{center}
\includegraphics[scale=0.35]{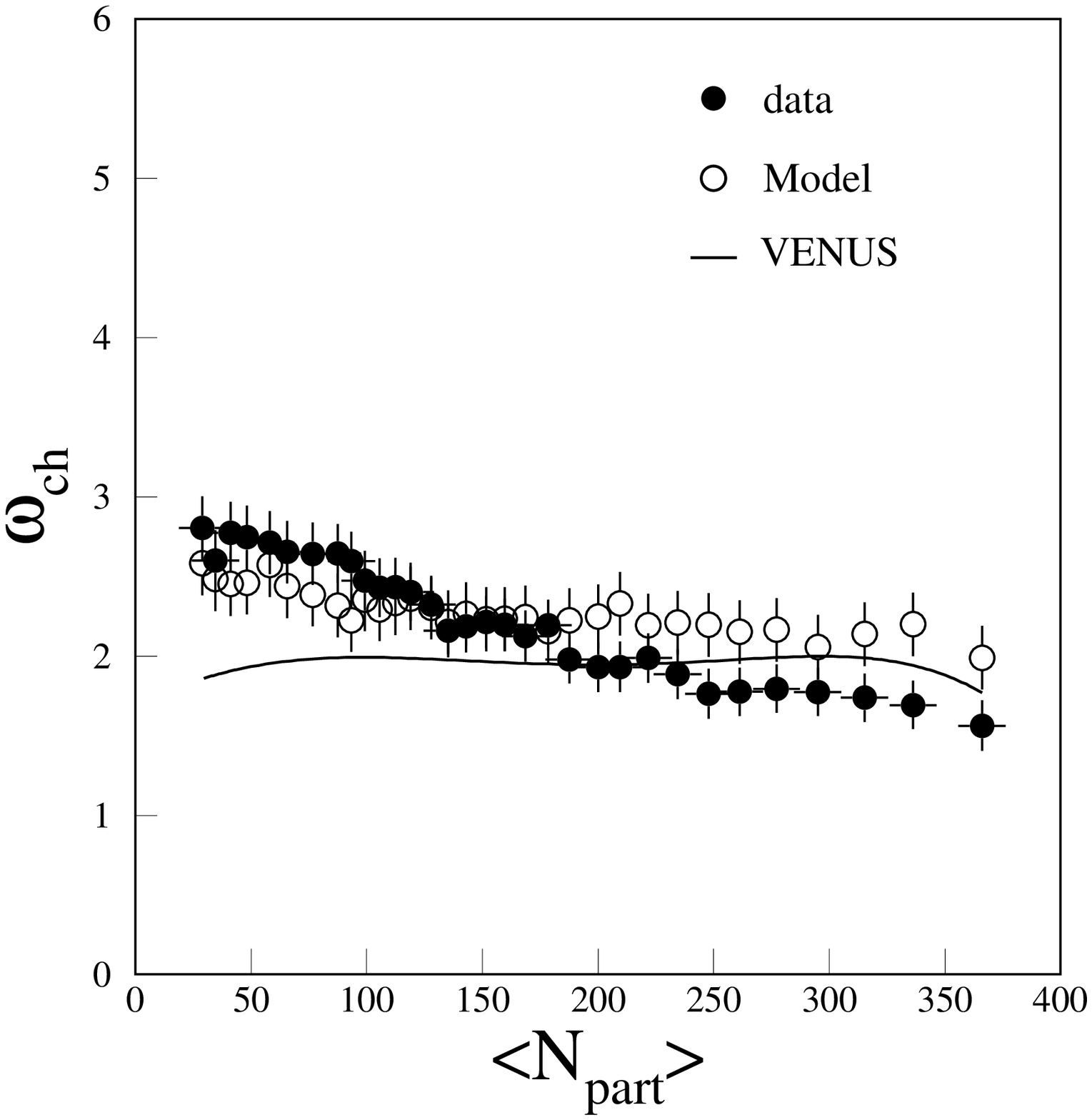}
\includegraphics[scale=0.35]{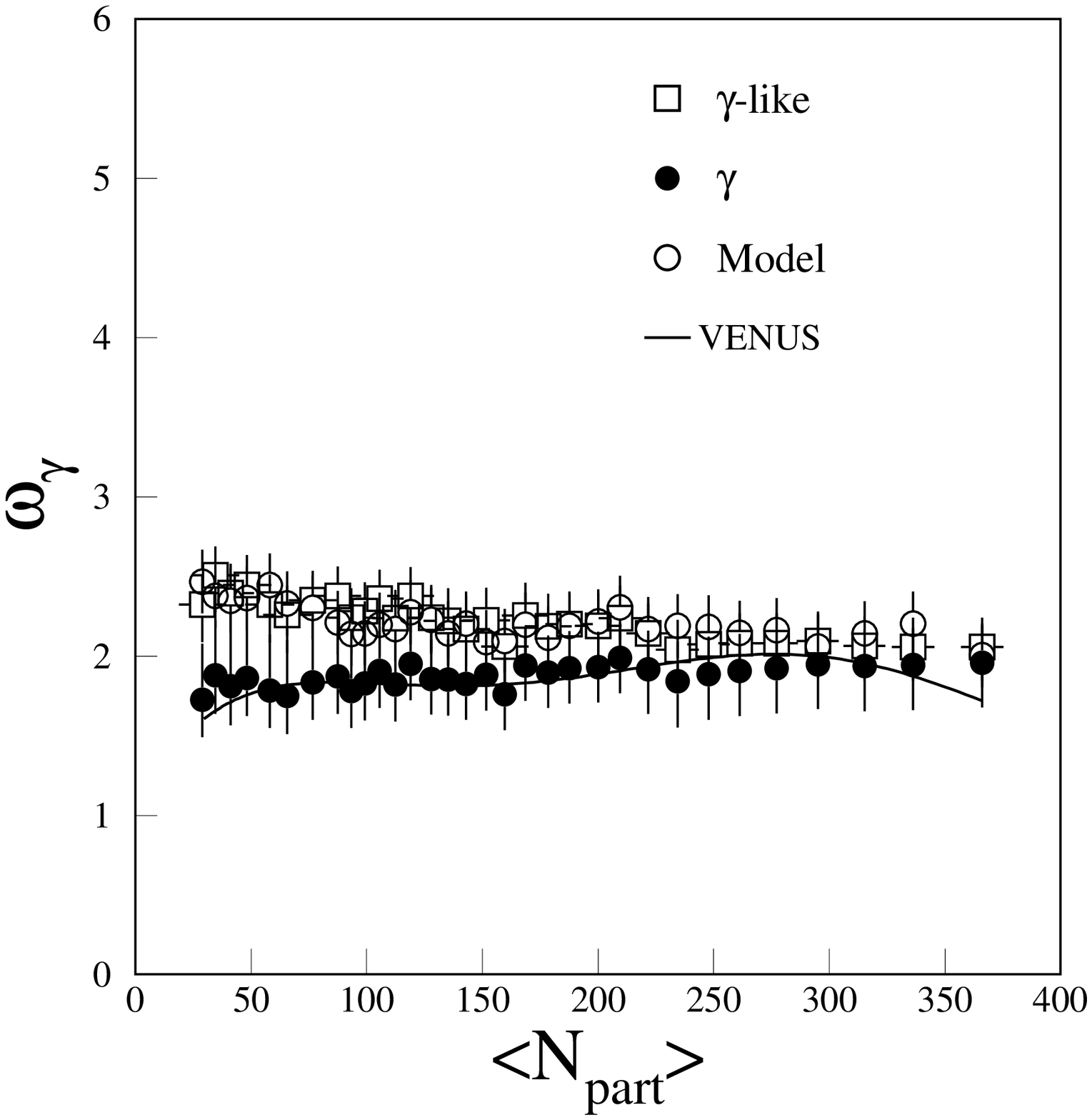}
\caption {\label{nch_ngam_w}
Multiplicity fluctuations of charged particles and photons for
various centralities in a region of common coverage of the PMD and SPMD as a 
function of the number of participants. 
The $\gamma-{\rm like}$ results correspond to fluctuations of the
measured photons without
correction for the photon counting efficiency or purity of the photon sample.
}
\end{center}
\end{figure}

It is very important to control the centrality selection carefully 
in fluctuation 
studies so that the impact parameter fluctuations are kept to a minimum and 
the distributions are good Gaussians. With this in 
mind we have used narrow centrality selections with bin widths of
2\% in cross section, as determined from the measured total 
transverse energy, such as, 
0-2\%, 2-4\%, 4-6\%, 6-8\%, etc. The fluctuations calculated 
for these centrality bins using Eqn.~1 are plotted in 
Fig.~\ref{nch_ngam_w} as a function of the number of participants. The results 
from data for both charged particles and photons are compared to those 
obtained from the VENUS event generator and a simple participant 
model~\cite{WA98-15}. In the participant model, where the total particle 
multiplicity in an event is the sum of the 
number of particles produced by each participant, the multiplicity
fluctuations, $\omega_{N}$, can be expressed as
\begin{equation}
       \omega_{N}  =  \omega_{n} + \langle n \rangle \omega_{N_{\mathrm part}}
\end{equation}
The value of $\omega_{n}$, the fluctuation in the number of particles 
falling within the detector acceptance ($n$) produced per participant 
can be obtained from nucleon-nucleon data.
The value of $\omega_{N_{\mathrm part}}$, the fluctuation in 
the number of participants, can be obtained from simulations.
From Fig.~\ref{nch_ngam_w} we observe that the measured fluctuations
are reasonably well reproduced by statistical models.

\subsection{Acceptance dependence of multiplicity fluctuations }

\begin{figure}
\begin{center}
\includegraphics[scale=0.35]{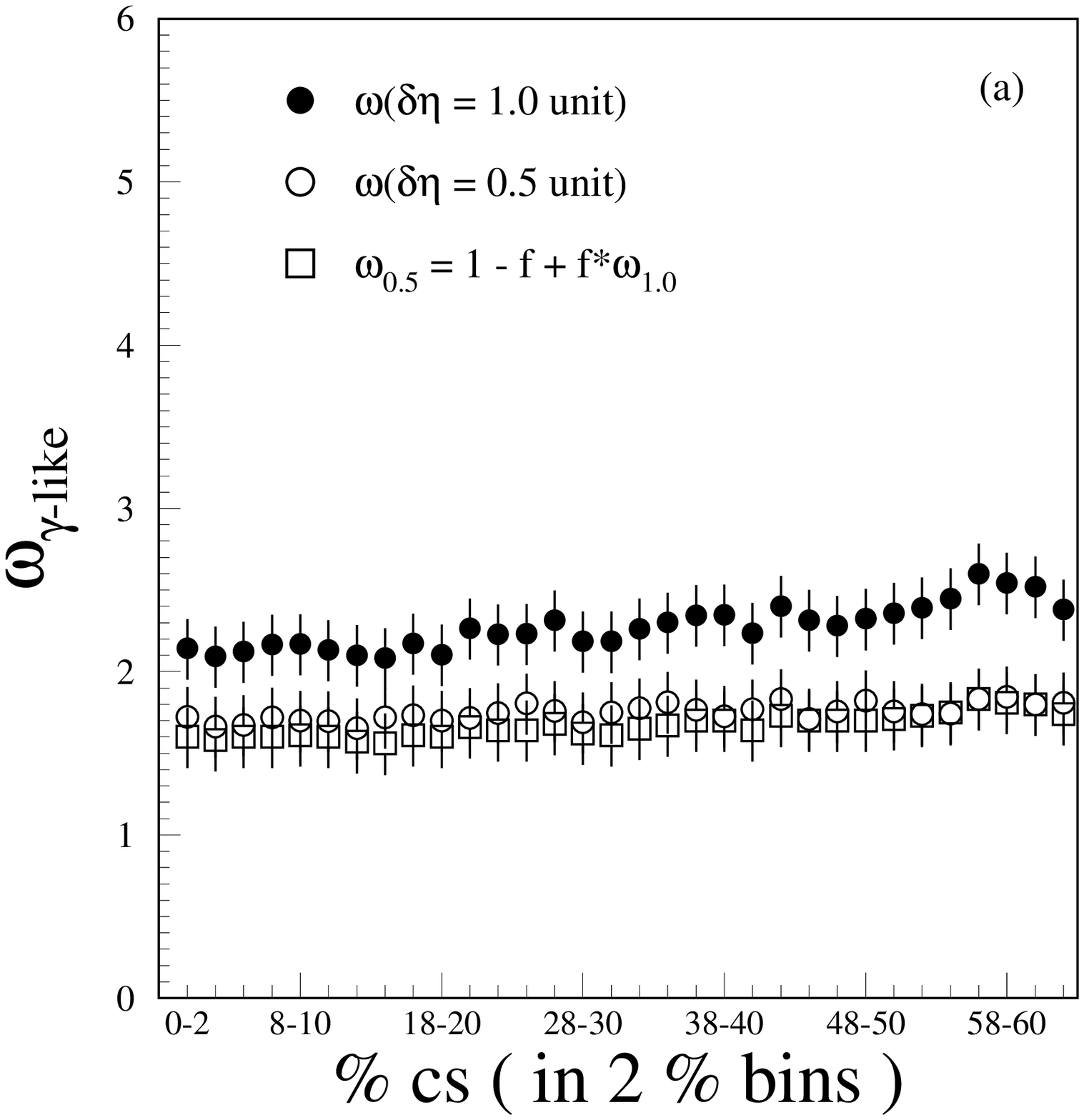}
\includegraphics[scale=0.35]{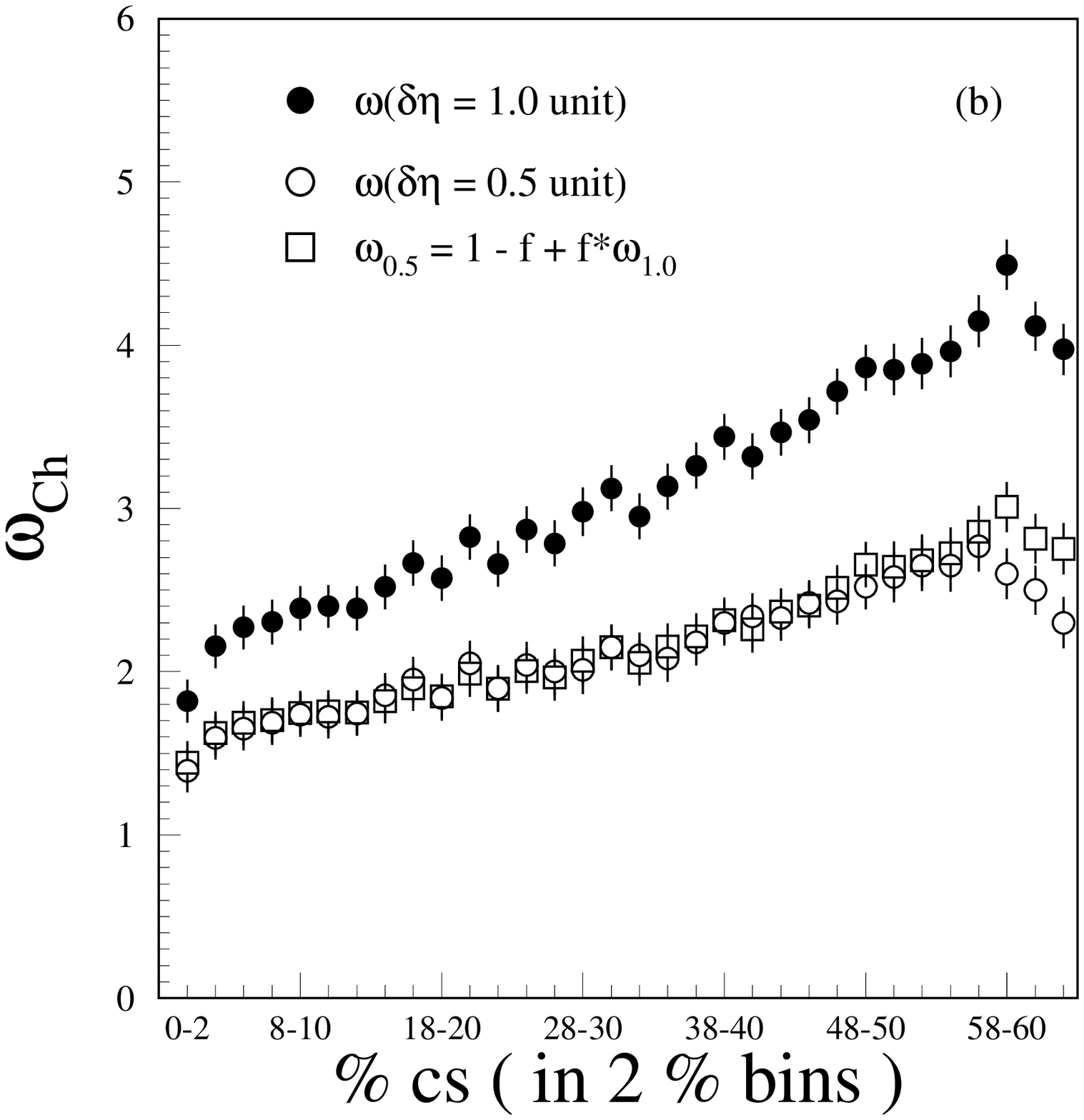}
\caption {\label{accp_w}
Multiplicity fluctuations
of photons and charged particles for two $\eta$ acceptance selections.
The open squares represent estimated fluctuation values
in $0.5$ unit of $\delta \eta$ from the observed fluctuations in $1.0$
unit of $\delta \eta$.
}
\end{center}
\end{figure}

Fluctuations in the particle multiplicity have also been studied by varying the 
rapidity acceptance. The results are shown in 
Fig.~\ref{accp_w}. One observes, contrary to naive expectation that 
the fluctuations should increase as the multiplicity decreases 
with decrease in 
acceptance, that instead the fluctuations decrease. 
This can be explained through a 
simple statistical model, based on the assumption that particles are accepted
following a binomial distribution. The details of this model are given 
in Ref~\cite{fluc_iop}. As per this model, the fluctuations
($\omega_n$ ) 
in a small acceptance region are related to the fluctuations ($\omega_m$) in a
larger acceptance as
\begin{equation}
\omega_n = 1 - f + f\omega_m
\end{equation}
where $f$ is the ratio of the average number of particles in the smaller
acceptance to the average number of particles in the larger acceptance.
The results from the model shown in Fig.~\ref{accp_w} are found to be in 
reasonably good agreement with the data.

\section{$N_{\gamma}$ vs. $N_{ch}$ fluctuations }

The main motivation to study event-by-event fluctuations in the photon vs. 
charged particle multiplicity is to search for possible formation of DCCs.
For this one would like a photon multiplicity detector and a charged particle 
multiplicity detector with as large a common coverage in $\eta-\phi$ 
phase space as possible. The main DCC search result from the SPS has come 
from the WA98 experiment~\cite{WA98-12,WA98-20}. 
First we discuss the results for top 
$5\%$ most central events. 
For the pseudo-rapidity and azimuthal region common to 
both PMD and SPMD used for this analysis, the average 
photon and charged particle 
multiplicity is 335 and 323, respectively. 
The experimental results are compared to simulated events and various types 
of mixed events to investigate possible DCC formation. 
The simulated events were generated by passing the VENUS output through a 
detector simulation package (GEANT), which incorporates the WA98 
experimental setup.  Four different
types of mixed 
events were generated from the data 
with equivalent event sample sizes as for real events. 
The details of the construction of the mixed events can be found in 
Ref.~\cite{WA98-12}. In Table~1 we summarize the construction and the 
physics issue probed by each type of mixed event.
\begin{table}
\begin{center}
\caption{Types of mixed events and the types of fluctuations probed.}
\smallskip
\begin{tabular}{|cccc|} \hline
             & PMD      & SPMD       & Type of fluctuation probed: \\
\hline
M1           & Mix hits & Mix hits   & Correlated + Individual \\

M2           & Unaltered hits &  Unaltered hits   & Correlated \\  
M3-$\gamma$  &  Unaltered hits &  Mix hits  & $N_{\gamma}$ only \\
M3-ch        &  Mix hits &    Unaltered hits  & $N_{\rm ch}$ only \\
\hline
\end{tabular}
\end{center}
\end{table}
A common analysis was carried out on the data, simulated, and mixed
events to investigate the source of any observed fluctuations. 

\subsection{$N_{\gamma}$ vs. $N_{ch}$ correlation analysis }

\begin{figure}
\begin{center}
\includegraphics[scale=0.4]{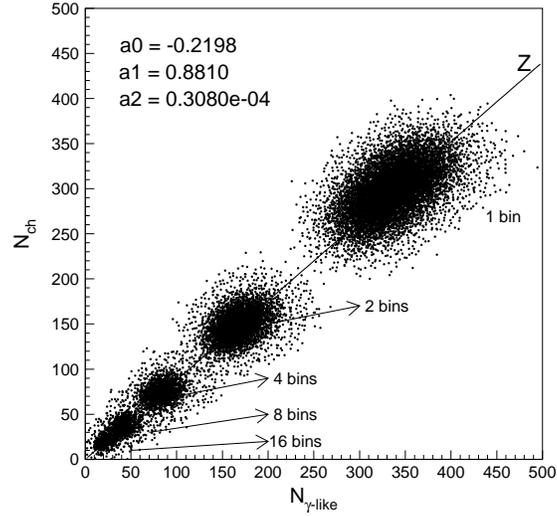}
\caption{ The event-by-event correlation between $N_{\mathrm {ch}}$ and
$N_{\gamma-{\mathrm {like}}}$ for the top $5\%$ centrality class. 
Overlaid on the plot is the common correlation axis (Z-axis). 
} 
\label{corr}
\end{center}
\end{figure}

\begin{figure}
\begin{center}
\includegraphics[scale=0.35]{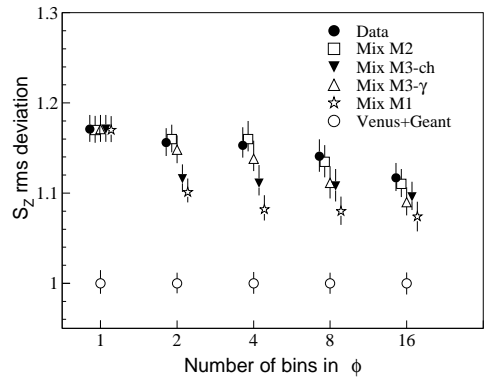}
\caption {\label{sz_rms}
The root mean square (RMS) deviations of the $S_Z$ distribution
for various divisions in the azimuthal angle. Errors include both
statistical and systematic sources.
}
\end{center}
\end{figure}

Analysis of the correlation between $N_{\gamma}$ and
$N_{\rm ch}$ is useful to search for DCC-type fluctuations.
The details of this can be found in Ref.~\cite{WA98-3,WA98-12}. From the 
event-by-event correlation between $N_{\gamma-{\rm like}}$ and $N_{\rm ch}$ 
in various $\phi$-segments (obtained by dividing the $\phi$-space into 
2, 4, 8, and 16 bins) and a common correlation axis ($Z$), one 
can obtain the closest distance ($D_{Z}$) of the data points to the 
correlation axis. The correlation plot is shown in Fig.~\ref{corr}.
In order to compare the fluctuations for different $\phi$ bins on a 
similar footing, a scaled variable, $S_{Z} = D_Z/s(D_Z)$, is used, 
where $s(D_Z)$ represents the RMS deviation of the $D_{Z}$ distribution 
for VENUS events analyzed in the same manner. 
The width of the distribution of $S_Z$ represents the
relative fluctuations of $N_{\gamma-{\rm like}}$ and $N_{\rm ch}$
from the correlation axis at any given $\phi$ bin.
Since the width of the $S_Z$ distribution quantifies the amount of
fluctuation, the RMS deviations of these distributions for data are
compared with 
those from mixed events and simulations in Fig.~\ref{sz_rms}.

The RMS deviations of the $S_Z$ distributions for mixed events and data 
agree for 1 bin  in $\phi$ (by construction) and for 16 bins in $\phi$ 
within the quoted errors.
However, for the other bins in $\phi$ one observes that results from
M1 mixed events are lower than those from data. This indicates the 
presence of localised non-statistical fluctuations in the data. The
source of the additional fluctuations is understood by 
the comparison to the M3-type of
mixed events. Comparison shows that the excess fluctuations have 
contributions both
from $N_{\gamma}$ and $N_{\mathrm ch}$. However the RMS deviations of
the M2-type of mixed events closely follow those from data, suggesting the
absence of event-by-event 
correlated DCC-like $N_{\gamma}$ vs. $N_{\mathrm ch}$ fluctuations.

\subsection{Multi-resolution analysis based on discrete wavelet 
transformation }

A multi-resolution analysis using discrete wavelet transformations 
(DWT) has been shown to be quite powerful in the search for 
localized domains of DCC \cite{dccstr}. For the present 
DWT analysis the full azimuthal region is divided into smaller bins 
in $\phi$, the number of bins at a given scale $j$ being $2^j$. The input 
to the analysis is a spectrum of the sample function at the smallest bin in
$\phi$ corresponding to the highest resolution scale, $j_{max}$(= 5 here). 
In the 
present case, the sample function is chosen to be the photon fraction, 
given as,
\begin{equation}
f^\prime(\phi) = {N_{\gamma-{\mathrm {like}}}(\phi)}/
{(N_{\gamma-{\mathrm {like}}}(\phi)+N_{\mathrm {ch}}(\phi))}
\end{equation}
The output of the DWT consists of a set of wavelet or father function
coefficients (FFC) at each scale, from $j=1$,...,($j_{max}-1$).
The coefficients obtained at a given scale, $j$, are derived from 
the distribution of the sample function at one higher scale, $j+1$.
The FFCs quantify the deviation of the bin-to-bin fluctuations in
the sample function at that higher scale relative to the average
behavior. The presence of localized non-statistical fluctuations
will increase the RMS deviation of the distribution of FFCs and may 
result in non-Gaussian tails \cite{dccstr}. 
Once again, comparing the RMS deviations of the FFC distributions of data, 
mixed events, and VENUS events may allow to draw inference about
the presence of localized fluctuations.
\begin{figure}
\begin{center}
\includegraphics[scale=0.35]{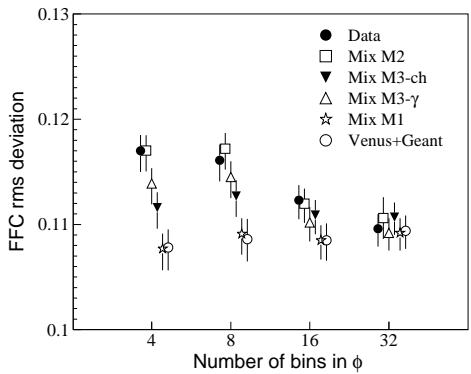}
\caption {\label{ffc_rms}
The root mean square (RMS) deviations of the FFC distribution
for various divisions in the azimuthal angle. Errors are due to both
statistical and systematic sources.
}
\end{center}
\end{figure}

The RMS deviations of these FFC distributions are summarized in 
Fig.~\ref{ffc_rms}. The RMS deviations of the FFC distributions for 
the data, VENUS, and mixed events are found to be close to each other 
(within quoted errors) for the case of $32$ bins in $\phi$.
While the values for M2 mixed events are found 
to closely follow those of the data for all bins in $\phi$,
the RMS deviations for the M3 mixed events lie
between those of the data and M1 mixed events. These results are consistent 
with those obtained from the analysis of the $S_{Z}$ distributions. 
These observations indicate the absence of event-by-event localized correlated
fluctuations (DCC-like) between $N_{\gamma-{\mathrm {like}}}$ and 
$N_{\mathrm {ch}}$. However, they do indicate the 
presence of localized independent fluctuations 
in both photon and charged particle multiplicities for intermediate bin 
sizes in azimuth.

\subsection{Centrality dependence of $N_{\gamma}$ vs. $N_{ch}$ fluctuations }

\begin{figure}
\begin{center}
\includegraphics[scale=0.35]{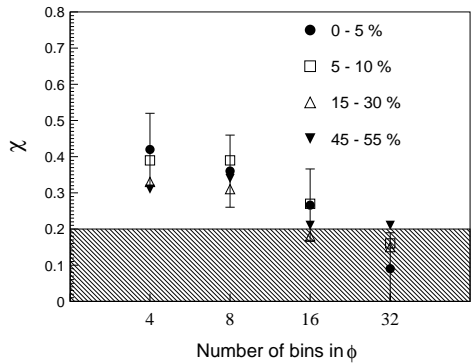}
\caption {\label{xi}
The fluctuation strength parameter  
for the four centrality classes. The error bars are
shown only for the top $5\%$ centrality class for clarity of presentation. 
The errors are similar for the other centralities.
}
\end{center}
\end{figure}

It is interesting to study the centrality dependence of 
$N_{\gamma}$ vs. $N_{ch}$ fluctuations~\cite{notsocentral}. 
This study has been carried 
out for four centrality classes corresponding to the top $5\%$, 
$5\%$ - $10\%$, 
$15\%$ - $30\%$, and $45\% - 55\%$ of the 
minimum bias cross section. The correlation and DWT analysis was carried out
on events from 
each of the centrality classes and the results in terms of RMS deviations
of $S_Z$ and FFC distributions for data, various sets of mixed events, and 
simulations were compared. For all four classes, the RMS deviations of
the $S_Z$ and FFC distributions of data and M2 mixed events agreed reasonably 
well with each other. However, comparison to M1 and M3 mixed events showed
the presence of localised uncorrelated event-by-event 
non-statistical fluctuations in both photon and charged particle 
multiplicities.
In order to quantify the strength of the total localised  
$N_{\gamma-{\mathrm {like}}}$ and $N_{\mathrm {ch}}$ fluctuations
for various bins in $\phi$ and for different centrality classes,
we define a sensitivity parameter $\chi$ as :
\begin{equation}
    \chi =  \frac{\sqrt{(s^2 - s_1^2)}}{s_1}
\label{chi_eqn}
\end{equation}
where $s_1$ and $s$ correspond to the RMS deviations
of the FFC distributions of the M1 mixed events and
real data, respectively. The results are shown in Fig.~\ref{xi} 
as a function of the number of bins in $\phi$ for the four different 
centrality classes. Qualitatively similar results are obtained when $\chi$ is 
calculated using the RMS deviations of the $S_{Z}$ distributions. 
The shaded portion indicates the region of $\chi$ where $s$ is
one $\sigma$ greater than the  RMS deviation FFC distributions for M1 events, 
where $\sigma$ is the total error on the M1 event RMS deviation. It
represents the limit above which a signal is detectable.
The result shows that the strength of the fluctuations 
decreases as the number of bins in $\phi$ increases, with a strength 
which decreases to below detectable level (within the quoted errors) 
for 16 and 32 bins. It is also observed that the strength of the signal 
decreases with decreasing centrality for 4 and 8 bins in azimuthal angle, 
although the tendency is not very strong. 

\subsection{Upper Limit on DCC production at SPS }

\begin{figure}
\begin{center}
\includegraphics[scale=0.35]{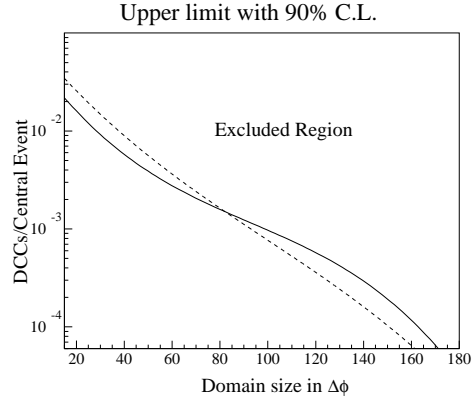}
\caption {\label{upper_limit}
The $90\%$ confidence level upper limit on DCC production for
central Pb+Pb collision at 158  A GeV/c 
as a function of the DCC domain size in azimuthal angle.
The solid line corresponds to data from the top $5\%$  and dashed line
to top $5-10\%$ of the minimum bias
cross section as determined by selection on the measured transverse energy
distribution. 
}
\end{center}
\end{figure}

If the amount of DCC-like fluctuations in the experimental data were
large, then the RMS deviations of $S_Z$ and FFC distributions 
for data would have been larger than those of M2
events. Since this is not the case, we have extracted upper limits on
the probability of DCC-like fluctuations at the 90\% confidence level
following the standard procedure as discussed in Ref.~\cite{WA98-20,uplim}. 
This is done within the context of the simple  
DCC model described in Ref.~\cite{WA98-20}. We give the upper limits for 
the two most central event classes ($0-5\%$ and $5-10\%$).
The $90\%$ CL upper limit contour has been calculated as 
$\chi$ + 1.28$e_{\chi}$, where $\chi$ is calculated 
using  Eqn.(\ref{chi_eqn}). Here $s_1$ and $s$ correspond to the RMS deviations
of the FFC (or $S_{Z}$) distributions for M2 mixed events and
real data, respectively, and $e_{\chi}$ is the error in $\chi$ from
the FFC (or $S_{Z}$) analysis. 
To relate the measured upper limit on the size of the fluctuations
in terms of DCC domain size and frequency
of occurrence we take use the simple 
simulated DCC model described in Ref.~\cite{WA98-20}.
The upper limit is set at that 
value of frequency of occurrence for a fixed DCC domain size at which
the $\chi$ value from the DCC model matches with that of the 
$\chi$ + 1.28$e_{\chi}$ upper limit from the experimental data. 
The  results for centrality classes 1 and 2 are shown in 
Fig.~\ref{upper_limit}. 

\section{Summary }

A detailed study of centrality and rapidity acceptance dependence of 
multiplicity fluctuations carried out by the 
WA98 experiment at the CERN SPS shows an absence of significant 
non-statistical fluctuations in the photon and charged particle multiplicity.
A model independent study of event-by-event correlated fluctuations (DCC-type) 
in photon and charged particle multiplicity using a robust mixed event 
technique reveals an absence of significant DCC-like correlated 
fluctuations at the SPS. 
However, the analysis indicates the presence of event-by-event uncorrelated 
fluctuations in both $N_{\gamma}$ and $N_{\mathrm {ch}}$
beyond those observed in simulated and  mixed events for limited region of 
azimuthal angle,
with the strength of fluctuation increasing
with increase in
centrality. Using the results from the data, mixed events,
and a simple model of DCC formation, an upper limit on DCC 
production in  Pb+Pb collisions at SPS energies has been set.
\paragraph{Acknowledgments}
One of us (B.M) is grateful to the Board of Research
on Nuclear Science and Department of Atomic Energy, 
Government of India for financial support in form of 
the Dr. K.S. Krishnan fellowship.
This work was supported jointly by
the German BMBF and DFG,
the U.S. DOE,
the Swedish NFR and FRN,
the Dutch Stichting FOM,
the Polish KBN 
the Grant Agency of the Czech Republic 
the DAE, DST, CSIR, UGC of the Government of India,
the PPE division of CERN,
the Swiss National Fund,
The Ministry of Education, Science and Culture,
the University of Tsukuba Special Research Projects, and
the JSPS Research Fellowships for Young Scientists.

\end{document}